\documentclass[showkeys,showpacs,amssymb,amsmath,preprint]{revtex4}
\usepackage{amsmath}
\usepackage{amssymb}
\usepackage[dvips]{graphicx}
\usepackage{dcolumn}
\usepackage{bm}
\usepackage[colorlinks=false,dvipdfm]{hyperref} 
\usepackage{setspace}

\begin{document}

\title{Local correlations of mixed two-qubit states \footnote{Physics Letters A, 374 (2010) 2429-2433.
\\http://dx.doi.org/10.1016/j.physleta.2010.04.004} }

\author{Fu-Lin Zhang}
\affiliation{Theoretical Physics Division, Chern Institute of
Mathematics, Nankai University, Tianjin, 300071, P.R.China \\PHONE:
011+8622-2350-9287, FAX: 011+8622-2350-1532}

\author{Chang-Liang Ren}
\affiliation{Department of Physics, Korea University, Seoul 136-713,
Korea}
\affiliation{Department of Modern Physics, University of
Science and Technology of China, Hefei, Anhui 230026, P.R.China}

\author{Ming-Jun Shi}
\affiliation{Department of Modern Physics, University of Science and
Technology of China, Hefei, Anhui 230026, P.R.China}

\author{Jing-Ling Chen}
\email[Email:]{chenjl@nankai.edu.cn}\affiliation{Theoretical Physics
Division, Chern Institute of Mathematics, Nankai University,
Tianjin, 300071, P.R.China \\PHONE: 011+8622-2350-9287, FAX:
011+8622-2350-1532}

\date{\today}

\begin{abstract}
The quantum probability distribution arising from single-copy von
Neumann measurements on an arbitrary two-qubit state is decomposed
into the local and nonlocal parts, in the approach of Elitzur,
Popescu and Rohrlich [A. Elitzur, S. Popescu, and D. Rohrlich, Phys.
Lett. A 162, 25 (1992)]. A lower bound of the local weight is proved
being connected with the concurrence of the state $p_L^{\max}=
1-\mathcal{C}(\rho)$. The local probability distributions for two
families of mixed states are constructed independently, which accord
with the lower bound.
\end{abstract}



\maketitle



\section{Introduction}

Entanglement and nonlocality are two fundamental concepts in quantum
description of nature, which are closely interconnected but not
identical \cite{EPR,Bell,Werner1989}. The former depicts the
nonseparability
 of the state of a composite quantum system \cite{EPR}, while the
latter is characterized by violation of a Bell inequality
\cite{Bell}, which means the local measurement outcomes of the state
cannot be described by a local hidden variables (LHV) model. It has
been proved that all pure entangled states violate such an
inequality and, consequently, are nonlocal \cite{Gisin1991}.
But Werner \cite{Werner1989}
has shown a family of mixed entangled states (called Werner states
now) can be described by a LHV model. The two concepts are not only
the fundamental features of quantum theory, but also the crucial
resources in quantum information \cite{Book,code,key,key2}.

To quantify the degree to which a state is entangled, several
measures have been proposed, such as entanglement of formation
\cite{EOF,Wootters97,Wootters98}, entanglement of distillation
\cite{EOD}, relative entropy of entanglement \cite{REE}, negativity
\cite{NEG,NEG1}, and so on. For two-qubit systems, the entanglement
of formation is equivalent to a computable quantity, which is
referred to as \textit{concurrence} \cite{Wootters97,Wootters98}.
The concurrence of a pure two-qubit state $| \psi \rangle= c_1 | 00
\rangle +c_2 | 01 \rangle  +c_3 | 10 \rangle  +c_4 | 11 \rangle  $
is given by
\begin{eqnarray} \label{ConP}
\mathcal{C}(| \psi \rangle)=2 |c_1 c_4 - c_2 c_3|.
\end{eqnarray}
The pure state is equivalent to
\begin{eqnarray}\label{PureTh}
| \psi (\theta) \rangle = \cos \theta | 00 \rangle + \sin \theta |
11 \rangle,\ \ \ \theta \in [0,\pi/4],
\end{eqnarray}
under local unitary (LU) transformations \cite{Book}, with
concurrence $\mathcal{C}(| \psi (\theta) \rangle)=2 \cos \theta \sin
\theta = \sin 2\theta$. For a mixed state, the concurrence is
defined as the average concurrence of the pure states of the
decomposition, minimized over all decompositions of $\rho =
\sum_{j}p_{j} | \psi_j \rangle \langle \psi_j |$,
\begin{eqnarray}\label{ConM}
\mathcal{C}(\rho)= \min \sum_{j}p_{j}\mathcal{C}(| \psi_j \rangle).
\end{eqnarray}
It can be expressed explicitly as \cite{Wootters97,Wootters98}
\begin{eqnarray}\label{ConEx}
\mathcal{C}(\rho)= \max
\biggr\{0,\sqrt{\lambda_{1}}-\sqrt{\lambda_{2}}-\sqrt{\lambda_{3}}-\sqrt{\lambda_{4}}
\biggr\},
\end{eqnarray}
in which $\lambda_{1},...,\lambda_{4}$ are the eigenvalues of the
operator $R=\rho (\sigma_{y} \otimes \sigma_{y} ) \rho^{*}
(\sigma_{y} \otimes \sigma_{y} )$ in decreasing order and
$\sigma_{y}$ is the second Pauli matrix.


The correlation in a bipartite quantum system is characterized by
the probability distribution
$P_{Q}(\alpha,\beta|\mathbf{a},\mathbf{b})$ of the outcomes $\alpha$
and $\beta$, corresponding to the measurements labeled by
$\mathbf{a}$ and $\mathbf{b}$ on the two subsystems respectively. It
is called \textit{local}, if the probability distribution can be
simulated by a LHV model. Namely, there exists a shared classical
variable $\lambda$ distributed with probability measure $\mu$ such
that
\begin{eqnarray}\label{LHV}
P_{Q}(\alpha,\beta|\mathbf{a},\mathbf{b}) = \int d\mu(\lambda)
P(\alpha|\mathbf{a},\lambda)P(\beta|\mathbf{b},\lambda),
\end{eqnarray}
where $P(\alpha|\mathbf{a},\lambda)$ and
$P(\beta|\mathbf{b},\lambda)$ are the local response functions of
the two observers. The form of the distribution in Eq. (\ref{LHV})
leads to a set of constraints on the local correlation (Bell-type
inequalities), for any fixed number of measurements on each
subsystem. Therefore, Bell inequality violation is a sufficient
condition of \textit{quantum nonlocality}.




Elitzur, Popescu, and Rohrlich (EPR2) \cite{EPR2} discuss the local
and nonlocal contents of nonlocal probability distributions [see Eq.
(\ref{decom})] from a different point of view. Actually, EPR2
approach can be abstractly interpreted to answer such a question:
whether an alternative description of nature is valid. Since the
original work of EPR2 appeared, few papers generalized it in depth.
Recently, as the approach is related to a more noticeable question,
the simulation of quantum correlations with other resource, it
attracted someone's attention again. Barrett \emph{et. al} gave an
upper bound of the weight of local component in $d\times d$ system
\cite{Barrett2006}. In his recent work \cite{Scarani2008}, Scarani
reviewed the previous results and decomposed the
 quantum correlation $P_Q$ corresponding to von Neumann measurements
 performed on the pure state (\ref{PureTh}) into a mixture of a
 local correlation $P_L$ and a nonlocal correlation $P_{NL}$
 \begin{eqnarray}\label{decom}
 P_Q=p_L(\theta) P_L + [1-p_L(\theta)] P_{NL},
 \end{eqnarray}
in EPR2 approach.
Scarani's construction of the local probability distribution $P_L$
leads to
\begin{eqnarray}\label{BoundP}
p_L(\theta)=1-\sin 2\theta,
\end{eqnarray}
which is an improved lower bound of $p_L^{\max}(\theta)$ on the
original result $p_L^{\max}(\theta)\geq (1-\sin 2\theta)/4$ given by
EPR2. Here, $p_L^{\max}(\theta)$ denotes the maximum weight of the
local component in Eq. (\ref{decom}). Further more, he presented an
upper bound for $p_L^{\max}(\theta)$ on the family of pure two-qubit
states and the first example of a lower bound on the local content
of pure two-qutrit states.

It is interesting to note that the proportion of nonlocal
correlation $P_{NL}$ in Scarani's construction is nothing but the
concurrence of $| \psi (\theta) \rangle$, $1-p_L(\theta)=\sin
2\theta$. The main aim of this paper is to show this result can be
generalized straightway to the mixed states case. Namely, we present
a construction of $P_L$ for arbitrary states $\rho$ of two qubits,
corresponding to the local weight $p_L(\rho)=1-\mathcal{C}(\rho)$.
The construction will be proved as a theorem in Sec. \ref{Mixed}. In
addition, we will give the EPR2 decompositions of some typical
states in quantum information, such as the Generalized Werner state
\cite{RhoWG} and the mixture of a Bell state and a mixed diagonal
state, of which Werner state and maximally entangled mixed states
$\rho_{MEMS}$ \cite{MEMS} are two spacial cases. Conclusion will be
made in the last section.




\section{EPR2 decompositions of Mixed Two-Qubit states}\label{Mixed}

\subsection{General Results}
The probality that the local von Neumann measurements labeled by
unit vectors $\mathbf{a}$ and $\mathbf{b}$ performed on the two
qubits with state $\rho$ lead to the outcomes ($\alpha$, $\beta$) is
\begin{eqnarray}\label{Pq}
P_{Q} (\alpha,\beta|\rho;\mathbf{a},\mathbf{b}) = \mathrm{Tr}
(\Pi_{A}\otimes \Pi_{B}  \rho),
\end{eqnarray}
with $\alpha,\beta=\pm 1$. Here, the projectors are given by
\begin{eqnarray} \label{Pro}
&&\Pi_{A} =
\frac{1}{2}(\mathbf{1}+\vec{\mathbf{\sigma}}\cdot\mathbf{A}),\nonumber
\\
&&\Pi_{B} =
\frac{1}{2}(\mathbf{1}+\vec{\mathbf{\sigma}}\cdot\mathbf{B}),
\end{eqnarray}
where $\mathbf{1}$ is the $2 \times 2$ unit matrix,
$\vec{\sigma}=(\sigma_x, \sigma_y, \sigma_z )$ are the Pauli
matrices in vector notation, and $\mathbf{A}=\alpha \mathbf{a}$ and
$\mathbf{B}=\alpha \mathbf{b}$ are unit vectors. Then, the quantum
probability distribution of the pure state $|\psi(\theta) \rangle$
can be obtained easily
\begin{eqnarray}\label{PqP}
P_{Q} (\theta) =\frac{1}{4}[1+c(A_z+B_z)+A_z B_z +s(A_x B_x - A_y
B_y )],\ \
\end{eqnarray}
where $c=\cos 2\theta$ and $s=\sin 2\theta$ as denoted in
\cite{Scarani2008}. Scarani improved the local probability
distribution on EPR2's original construction
to
\begin{eqnarray}\label{PlP}
P_{L}=\frac{1}{4}[1+f(A_z)][1+f(B_z)],
\end{eqnarray}
with the function $f(x)=\mathrm{sgn}(x)\min(1,\frac{c}{1-s}|x|)$.
This keeps the product form in \cite{EPR2} and leads to
$p_L=1-s=1-\mathcal{C}(|\psi(\theta)\rangle)$.

Whereas, the product form construction of $P_L$ is obviously not
optimal for mixed states because of the presence of classical
correlation. A simple example is the separable state
$\rho_s=(|00\rangle \langle 00|+|11\rangle \langle 11|)/2$, whose
quantum probability distribution $P_Q(\rho_s)=(1+A_z B_z)/4$ should
be completely local. One can easily find the following equation is
self-contradictory,
\begin{eqnarray}
P_Q(\rho_s)=\frac{1}{4}(1+f_A)(1+f_B), \ \ f_A, f_B \in [-1,1],
\end{eqnarray}
if $f_A$ and $f_B$ are requested to be odd functions of $\mathbf{A}$
and $\mathbf{B}$ respectively. Actually, a straightforward
construction of the local correlation of $\rho_s$ is
\begin{eqnarray}
P_L (\rho_s)&=&\frac{1}{2} F^{+}(A_z) F^{+}(B_z) +\frac{1}{2}
F^{-}(A_z) F^{-}(B_z) \nonumber\\
&=&P_Q(\rho_s),
\end{eqnarray}
with $F^{\pm}(x)=\frac{1}{2}(1 \pm x)$. It contains a two-outcomes
random variable with equiprobability as the LHV. The following
results will show the local weight of an arbitrary two-qubit state
satisfies $p_L= 1- \mathcal{C}(\rho)$, if we choose the local
probability distribution with a discrete LHV as
\begin{eqnarray}\label{PlM}
P_L= \sum_{i} \mu_{i} p_{i}(\mathbf{A})q_{i}(\mathbf{B}),
\end{eqnarray}
where $ \mu_{i}, p_{i}(\mathbf{A}), q_{i}(\mathbf{B}) \in [0,1]$ are
probabilities satisfying $\sum_{i} \mu_{i}=1$, $p_{i}(\mathbf{A}) +
p_{i}(\mathbf{-A})=1$ and $q_{i}(\mathbf{B}) +
q_{i}(\mathbf{-B})=1$.

\emph{Theorem 1.} The local content of the probability distribution
for a two-qubit state $\rho$ has a lower bound $p_L^{\max} (\rho)
\geq 1- \mathcal{C}(\rho)$.

\emph{Proof.} According to the procedure given by Wootters
\cite{Wootters98}, one can always obtain a decomposition $\{ |
\phi_i \rangle \}$
 minimizing the average concurrence in Eq. (\ref{ConM}), $\rho= \sum_i t_i
| \phi_i \rangle \langle \phi_i |$, in which $\sum_i t_i =1$ and all
the elements have the same value of concurrence as the mixed state
$\rho$.
The elements are equivalent under LU transformation to the same
state in the form of Eq. (\ref{PureTh})
\begin{eqnarray}
| \phi_i \rangle \ =U^A_i \otimes U^B_i | \psi(\theta) \rangle,
\end{eqnarray}
with the concurrence $\mathcal{C}(| \psi(\theta)
\rangle)=\mathcal{C}( \rho)$.

Denote the unit vectors by $\mathbf{A}^{(i)}$ and
$\mathbf{B}^{(i)}$, which satisfy
$\vec{\mathbf{\sigma}}\cdot\mathbf{A}^{(i)}= U^{A \dag}_i
\vec{\mathbf{\sigma}}\cdot\mathbf{A} U^A_i $ and
$\vec{\mathbf{\sigma}}\cdot\mathbf{B}^{(i)}= U^{B \dag}_i
\vec{\mathbf{\sigma}}\cdot\mathbf{B} U^B_i $. The quantum
probability distribution is straightforward to obtain
\begin{eqnarray}
P_Q (\rho) = \sum_i t_i P_Q^{(i)},
\end{eqnarray}
where $P_Q^{(i)}= \langle \psi(\theta)| \Pi^{(i)}_A \otimes
\Pi^{(i)}_B |\psi(\theta) \rangle$ with $\Pi^{(i)}_A =
\frac{1}{2}[\mathbf{1}+\vec{\mathbf{\sigma}}\cdot\mathbf{A}^{(i)}]$
and $\Pi^{(i)}_B =
\frac{1}{2}[\mathbf{1}+\vec{\mathbf{\sigma}}\cdot\mathbf{B}^{(i)}]$.
Each $P_Q^{(i)}$ can be decomposed in Scarani's approach as
\begin{eqnarray}
P_Q^{(i)}=[1-\mathcal{C}(\rho)] P_L^{(i)}+ \mathcal{C}(\rho)
P_{NL}^{(i)},
\end{eqnarray}
where $P_L^{(i)}$ is defined in the form of Eq. (\ref{PlP}) with
$\mathbf{A}^{(i)}$ being substituted for $\mathbf{A}$ and
$\mathbf{B}^{(i)}$ for $\mathbf{B}$. A natural construction of the
local probability distribution is $P_L (\rho) = \sum_i t_i
P_L^{(i)}$, taking the form in Eq. (\ref{PlM}). Then, one can obtain
\begin{eqnarray}
P_Q (\rho)=[1-\mathcal{C}(\rho)]P_L (\rho) + \mathcal{C}(\rho)\sum_i
t_i P_{NL}^{(i)},
\end{eqnarray}
which ends the proof. $\square$

Since the procedure given by Wootters \cite{Wootters98} to derive
the optimal decomposition in Eq. (\ref{ConM}) is effective but not
easy to implement, we give the EPR2 decompositions of two families
of typical mixed states in the following parts of this section.
These are constructed directly, independent of the process presented
above.

\subsection{Werner State \& Generalized Werner State}

The Werner state \cite{Werner1989} takes the form as
\begin{eqnarray} \label{rhoW}
\rho_{W} = x | \psi^{+} \rangle \langle  \psi^{+}| +(1-x)
\frac{\mathbf{1} \otimes \mathbf{1}}{4}, \ \ \ x\in [0,1],
\end{eqnarray}
where $| \psi^{+} \rangle= [| 00 \rangle + | 11 \rangle]/ \sqrt{2}$
is one of the Bell basis. The concurrence $\mathcal{C}(\rho_W)=\max
\{0, (3x-1)/2  \}$. And its quantum probability distribution is
given by
\begin{eqnarray}\label{PqW}
P_Q (\rho_W)=\frac{1}{4}[1+x(A_z B_z +A_x B_x -A_y B_y)].
\end{eqnarray}
When $x=1/3$, $\rho_{W}$ is separable, and $ P_Q (\rho_W)$ can be
represented as a local form
\begin{eqnarray}\label{PlW31}
P_L^{1/3}(\rho_W)
&=&\frac{1}{6}[F^{+}(A_z)F^{+}(B_z)+F^{-}(A_z)F^{-}(B_z)
+F^{+}(A_x)F^{+}(B_x)\nonumber\\
&\ & +F^{-}(A_x)F^{-}(B_x)
+F^{+}(A_y)F^{-}(B_y)+F^{-}(A_y)F^{+}(B_y)]\nonumber\\
&=&P_Q(\rho_W)|_{x=1/3}.
\end{eqnarray}
If we define the local distribution as
\begin{eqnarray}\label{PlW}
P_L(\rho_W) = \left \{
\begin{array}{lr}
P_L^{1/3}(\rho_W),     \; &  x \geq 1/3 \;;\\
3xP_L^{1/3}(\rho_W) + (1-3x)\frac{1}{4}, \; &  x<1/3 \;;
\end{array}
\right.
\end{eqnarray}
it is easy to prove $P_Q(\rho_W)=P_L(\rho_W)$ for $x<1/3$ and
$P_Q(\rho_W)/P_L(\rho_W) \geq \frac{3}{2}(1-x)$ for $x\geq1/3$. The
minimum of the radio occurs when the unit vectors $\mathbf{A} \cdot
\mathbf{B'}=-1$ with $ \mathbf{B'} = (B_x, -B_y, B_z)$.
 This
indicates the local content of Werner state
$p_{L}(\rho_W)=1-\mathcal{C}(\rho_W)$ corresponding to the
construction of $P_L(\rho_W)$ in Eq. (\ref{PlW}).

However, a better bound can be obtained easily based on the fact
that an entangled Werner state may admit a LHV model. In the seminal
work of Werner \cite{Werner1989}, he constructed a LHV model of the
 states (\ref{rhoW}) for $x \leq 1/2$ under von Neumann measurements.
This result has been extended to general measurements
\cite{WernerState2002} and more parties \cite{WernerState2006}. In
\cite{PhysRevA.73.062105}, Ac\'\i{}n \emph{et. al.} proved
the quantum probability distribution (\ref{PqW}) is local when the
parameter $x\leq x_c = 0.6595$ under von Neumann measurements.
Therefore, we can replace the demarcation point $1/3$ by $x_c$, and
define the separable function (\ref{PlW}) using $P_L^{x_c}(\rho_W)
=P_Q(\rho_W)|_{x=x_c}$ instead of $P_L^{1/3}(\rho_W)$. Choosing the
combinatorial coefficients in Eq. (\ref{PlW}) as $\{ x/x_c,1-x/x_c
\}$, we obtain a better bound
$p^{'}_{L}(\rho_W)=1-\mathcal{C}^{'}(\rho_W)$ with
$\mathcal{C}^{'}(\rho_W)=\max \{0, (x-x_c)/(1-x_c)  \}$. Whereas, it
is difficult to extended this result to any more general two-qubit
states. In the following paragraph, we will show the construction of
$P_L(\rho_W)$ in Eqs. (\ref{PlW31}) and (\ref{PlW}) can be
generalized to treat the states in Eq. (\ref{rhoGW}).

\begin{figure}
\centering
\includegraphics[width=8cm]{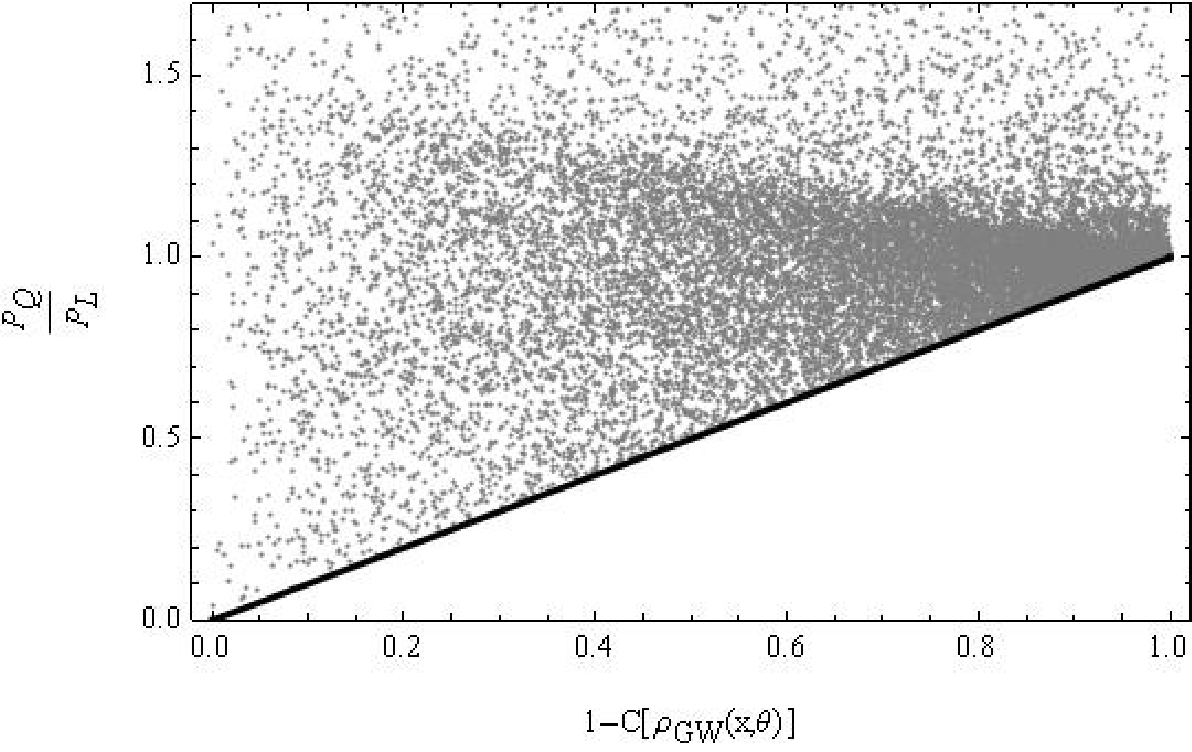} \\
 \caption{Plot of 20000 randomly generated sets of entangled $\rho_{GW}$ and unit vectors $(\mathbf{A},\mathbf{B
 })$ in the plane of $P_Q(\rho_{GW})/P_L(\rho_{GW}) \sim 1-\mathcal{C}(\rho_{GW})$ in company with the line of $P_Q(\rho_{GW})/P_L(\rho_{GW})= 1-\mathcal{C}(\rho_{GW})$.}
\label{fig1}
\end{figure}

A family of generalized Werner state \cite{RhoWG} is given by
\begin{eqnarray}\label{rhoGW}
\rho_{GW} = x | \psi(\theta) \rangle \langle  \psi(\theta) | +(1-x)
\frac{\mathbf{1} \otimes \mathbf{1}}{4},
\end{eqnarray}
which is the mixture of the pure state (\ref{PureTh}) with the
completely random state. Its concurrence is
$\mathcal{C}(\rho_{GW})=\max \{0, [(1+2s)x-1]/2 \}$,  and quantum
correlation can be obtained
\begin{eqnarray}\label{PqGW}
P_Q(\rho_{GW}) =\frac{1}{4}\{1+x[cA_z+cB_z+A_z B_z +s(A_x B_x - A_y
B_y)] \},
\end{eqnarray}
with $s$ and $c$ taking the definition in Eq. (\ref{PqP}). As the
treatment of Werner state, we start from the critical value of
$x_c=1/(1+2s)$, for which Eq. (\ref{PqGW}) is local obviously
\begin{eqnarray}\label{PlGWxc}
P_L^{x_c}(\rho_{GW})
&=&  \frac{x_c}{2}\{c_+ F^{+}(A_z)F^{+}(B_z)+c_- F^{-}(A_z)F^{-}(B_z)\nonumber\\
&& \ \ \ +sF^{+}(A_x)F^{+}(B_x)+sF^{-}(A_x)F^{-}(B_x)\nonumber\\
&& \ \ \ +sF^{+}(A_y)F^{-}(B_y)+sF^{-}(A_y)F^{+}(B_y)\}\nonumber\\
&=&P_Q(\rho_{GW})|_{x=x_c},
\end{eqnarray}
where $c_{\pm}=1\pm c$. When $x<x_c$, one can choose
\begin{eqnarray}\label{PlGWS}
P_L(\rho_{GW})&=& (1+2s)xP_L^{x_c}(\rho_{GW}) + [1-(1+2s)x]\frac{1}{4}\nonumber\\
&=&P_Q(\rho_{GW})
\end{eqnarray}
 For the entangled region $x>x_c$, an appropriate construction of
 local distribution is given by the linear combination
\begin{eqnarray}\label{PlGWE}
P_L(\rho_{GW})= k P_L + (1-k) P_L^{x_c}(\rho_{GW}),
\end{eqnarray}
where $P_L$ is the construction for pure state in Eq. (\ref{PlP}),
and $k=\frac{(1-s)[(1+2s)x-1]}{s[3-(1+2s)x]} \in[0,1]$ which is
derived from the equation
\begin{eqnarray}\label{Eqk}
\frac{c}{1- [(1+2s)x-1]/2} = k\frac{c}{1-s} + (1-k) \frac{c}{1+2s}.
\end{eqnarray}
Although we do not have a fully analytical proof, our numerical
evidence illustrates that the local probability distributions in Eq.
(\ref{PlGWE}) satisfies $P_Q(\rho_{GW})/P_L(\rho_{GW})\geq
1-\mathcal{C}(\rho_{GW})$. A detailed introduction is as follows: We
randomly generate one million sets of
$\{\theta,x,\mathbf{A},\mathbf{B} \}$, where the parameters satisfy
$x>x_c$ corresponding to entangled $\rho_{GW}$. Substituting them
into Eqs. (\ref{PqGW}) and (\ref{PlGWE}) and the concurrence of
$\rho_{GW}$, we find $P_Q(\rho_{GW})/P_L(\rho_{GW})\geq
1-\mathcal{C}(\rho_{GW})$ to come into existence. To show the
relation of inequality, in Fig. \ref{fig1}, we plot 20000 sets of
random $\{\theta,x,\mathbf{A},\mathbf{B} \}$ in the plane of
$P_Q(\rho_{GW})/P_L(\rho_{GW}) \sim  1-\mathcal{C}(\rho_{GW})$ in
company with the solid line of $P_Q(\rho_{GW})/P_L(\rho_{GW})=
1-\mathcal{C}(\rho_{GW})$. Consequently, choosing the local
probability distributions in Eqs. (\ref{PlGWS}) and (\ref{PlGWE}),
one has
\begin{eqnarray}
P_Q (\rho_{GW}) = [1-\mathcal{C}(\rho_{GW})]P_L
(\rho_{GW})+\mathcal{C}(\rho_{GW})P_{NL}(\rho_{GW}),
\end{eqnarray}
for arbitrary $\rho_{GW}$, \emph{i. e.} the local content
$p_L(\rho_{GW})= 1-\mathcal{C}(\rho_{GW})$.

\subsection{Mixture of a Bell State and a Diagonal State}
Another generalization of Werner state is the Bell State $| \psi^{+}
\rangle$ mixed with a diagonal state
\begin{eqnarray}\label{rhoBD}
\rho_{BD}=\begin{bmatrix}
 \ x+\gamma/2 & 0 & 0 & \gamma/2\\
 \ 0 & a & 0 & 0\\
 \ 0 & 0 & b & 0\\
 \ \gamma/2 & 0 & 0 & y+\gamma/2
\end{bmatrix},
\end{eqnarray}
where the non-negative real parameters $x+y+a+b+\gamma=1$. It
contains many special two-qubit states, such as maximally entangled
mixed states $\rho_{MEMS}$ \cite{MEMS}, frontier states of the
bounds for concurrence \cite{RhoFr} and so on. Its concurrence is
$\mathcal{C}(\rho_{BD})=\max \{0, \gamma-2\sqrt{ab} \}$, which is
independent on $x$ and $y$. Our construction of the EPR2
decomposition of $\rho_{BD}$ is divided into two steps:
(\romannumeral1) We give the results of the spacial case of $x=y=0$;
(\romannumeral2) The local distribution of the general case can be
derived immediately based on the results of the first step.

(\romannumeral1) When $x=y=0$, $\rho^0_{BD}=\gamma |\psi^+ \rangle
\langle \psi^+| +  a |01\rangle \langle 01| + b |10\rangle \langle
10| $, and the quantum probability distribution is
\begin{eqnarray}\label{PqBD0}
P_Q (\rho^0_{BD}) = \frac{1}{4}[1+ (a-b)(A_z - B_z) +(\gamma
-a-b)A_z B_z
 + \gamma (A_x B_x - A_y B_y)],\ \
\end{eqnarray}
where we choose $a\geq b$ without loss of generality.
At the critical point of separability $\gamma=2\sqrt{ab}$,
\begin{eqnarray}\label{PlBD0c}
P_L^{c}(\rho^0_{BD})&=&
\frac{1}{4}[F^{+}_A(A_x)F^{+}_B(B_x)+F^{-}_A(A_x)F^{-}_B(B_x)
+F^{+}_A(A_y)F^{-}_B(B_y)+F^{-}_A(A_y)F^{+}_B(B_y)] \nonumber
\\
&=&P_Q (\rho^0_{BD})|_{\gamma=2\sqrt{ab}}
\end{eqnarray}
where the local response functions $F^{\pm}_A(x)=\frac{1}{2}(1+ \sin
\vartheta A_z \pm \cos \vartheta x)$ and
$F^{\pm}_B(x)=\frac{1}{2}(1- \sin \vartheta B_z \pm \cos \vartheta
x)$, with $\vartheta = \vartheta_c= \arcsin (\sqrt{a}-\sqrt{b})$. In
the region $\gamma <  2\sqrt{ab}$, we assume
\begin{eqnarray}\label{PlBD0s}
P_L(\rho^0_{BD})&=& g P_L^{c}(\rho^0_{BD}) + (1-g)
P_L^{0}(\rho^0_{BD}), \\
P_L^{0}(\rho^0_{BD})&=& \lambda_+ F^{+}(A_z)F^{-}(B_z)+\lambda_-
F^{-}(A_z)F^{+}(B_z), \nonumber
\end{eqnarray}
where $\lambda_{\pm}=\frac{1}{2}(1\pm\Delta)$ and
$P_L^{0}(\rho^0_{BD})$ take the same form as $P_Q
(\rho^0_{BD})|_{\gamma=0}$. To hold the relation
$P_L(\rho^0_{BD})=P_Q(\rho^0_{BD})$ when $\gamma <  2\sqrt{ab}$, the
parameters should be chosen as $g=2\gamma /(\gamma+2\sqrt{ab})$ and
$\Delta=(\sqrt{a}+ \sqrt{b}-g)(\sqrt{a}- \sqrt{b}) /(1-g)$, both of
which lie in $[0,1]$. When $\rho^0_{BD}$ is entangled with $ \gamma
>  2\sqrt{ab}$, the probability distribution (\ref{PqBD0}) can be decomposed as
\begin{eqnarray}
P_Q (\rho^0_{BD}) = [1-\mathcal{C}(\rho^0_{BD})] P_L(\rho^0_{BD})+
\mathcal{C}(\rho^0_{BD})P_{NL}(\rho^0_{BD}),\ \
\end{eqnarray}
where $P_L(\rho^0_{BD})$ takes the definition in Eq. (\ref{PlBD0c})
with $\vartheta=\arcsin [(\sqrt{a}- \sqrt{b})/(\sqrt{a}+\sqrt{b})]$
and $P_{NL}(\rho^0_{BD})=\frac{1}{4}(1+\mathbf{A}\cdot \mathbf{B'})$
corresponding to the probability distribution of the Bell state
$|\psi^{+} \rangle $.

(\romannumeral2) An arbitrary state (\ref{rhoBD}) can always be
written as $\rho_{BD}=x|00\rangle \langle 00 | + y |11 \rangle
\langle 11 |+ p \rho'$, where $p=\gamma+a+b$ and
$\rho'=\rho^0_{BD}|_{(\gamma,a,b)\rightarrow(\gamma/p,a/p,b/p)}$.
Its concurrence is $\mathcal{C}(\rho_{BD})=p\mathcal{C}(\rho')$. One
can obtain immediately
\begin{eqnarray}
P_Q (\rho_{BD}) = p P_Q (\rho')+ x F^{+}(A_z) F^{+}(B_z) + y
F^{-}(A_z) F^{-}(B_z). \ \
\end{eqnarray}
In the approach given in step (\romannumeral1), $P_Q (\rho')$ can be
divided into the local $P_L(\rho')$ and nonlocal $P_{NL}(\rho')$
parts, with the wights $1-\mathcal{C}(\rho')$ and
$\mathcal{C}(\rho')$ respectively. Choosing the construction $P_L
(\rho_{BD}) = [p (1-\mathcal{C}(\rho'))P_L (\rho')+ x F^{+}(A_z)
F^{+}(B_z) + y F^{-}(A_z) F^{-}(B_z) ]/[p (1-\mathcal{C}(\rho'))+ x
 + y ]$, one has
\begin{eqnarray}
P_Q (\rho_{BD}) = [1-\mathcal{C}(\rho_{BD})]P_L
(\rho_{BD})+\mathcal{C}(\rho_{BD})P_{NL}(\rho'),
\end{eqnarray}
in which the nonlocal probability distribution is the same as the
one of $\rho'$.

\section{conclusion and discussion} \label{concl}

In conclusion, we investigate the EPR2 decomposition of the
probability distribution arising from single-copy von Neumann
measurements on arbitrary two-qubit states. In our constructive
proof, the local content is shown to have a lower bound connected
with the concurrence which measures the degree of entanglement,
$p_L^{\max} \geq 1-\mathcal{C}(\rho)$. The local probability
distribution for two families of mixed states are constructed
independent of the scheme in the proof. Both of them lead to the
local weight $p_L= 1-\mathcal{C}(\rho)$.

In this paper, what we concern about are the mixed states of a
two-qubit system. A natural extension of this issue is to study the
EPR2 decomposition in a bipartite arbitrary-dimensional system. To
our knowledge, only in Scarani's paper \cite{Scarani2008}, a
one-parameter family of two-qutrit states has been investigated in
the EPR2 approach. For the state $| \Psi (\gamma) \rangle= (| 00
\rangle+| 11 \rangle+ \gamma | 22 \rangle ) / \sqrt{2+\gamma^2}$,
Scarani chose the local distribution $P_L$ to be the product of
Kronecker deltas $\delta_{\alpha,i(A)}$ and $\delta_{\beta,j(B)}$,
where $i(A)$ and $j(B)$ are the most probable local outcomes when
$\gamma > 1$. His numerical results show the local content is
nonzero when $\gamma > 2$. However, an analytic lower bound of $p_L$
is absent. We would like to present our prospects to give an
improved lower bound and generalize it to the mixed states case.
(\romannumeral1) We start from the one-parameter state $| \Psi
(\gamma) \rangle$ and construct a local distribution $P_L$ which is
a function of the parameter $\gamma$. The Kronecker deltas
 can be represented as
$\delta_{\alpha,i(A)}=\frac{1}{3} \{ 1 + 2 \cos \frac{2 \pi}{3}[
\alpha - i (A)] \}$ and $\delta_{\beta,j(B)}=\frac{1}{3} \{ 1 + 2
\cos \frac{2 \pi}{3}[ \beta - j (B)] \}$, in which the cosine
functions play the roles of $\alpha \mathrm{sgn} (a_z)$ and  $\beta
\mathrm{sgn} (b_z)$ in the original construction of the qubit case
given by EPR2 \cite{EPR2}. To obtain an improved lower bound of
$p^{max}_L$, one can choose an appropriate function to substitute
for the cosine function, like Scarani introducing the function
$f(x)$ in Eq. (\ref{PlP}) to take the place of the sign function.
(\romannumeral2) A subsequent work is to extend the results of the
one-parameter state to the Schmidt-decomposed state $| \Phi
(\gamma_1,\gamma_2) \rangle= (| 00 \rangle+ \gamma_1 | 11 \rangle+
\gamma_2 | 22 \rangle ) / \sqrt{1+\gamma_1^2+\gamma_2^2}$.
Obviously, the lower bound of $p^{max}_L$ for $| \Phi
(\gamma_1,\gamma_2) \rangle$ should be a function of the parameters
$\gamma_1$ and $\gamma_2$, and afterward, be a function of the
entanglement invariants of the two-qutrit state \cite{Fei}.
(\romannumeral3) Based on the results in the first two steps, one
can attempt to decompose some typical mixed two-qutrit states in the
EPR2 approach. In the light of the experience in Sec. \ref{Mixed},
an alternative construction of the local distribution $P_L$ has the
form of a linear combination of the pure states case. And it is
often effective to start from the critical point of separability.





\begin{acknowledgments}
This work is supported in part by NSF of China (Grants No.
10975075), Program for New Century Excellent Talents in University,
and the Project-sponsored by SRF for ROCS, SEM.
\end{acknowledgments}

\bibliography{LocalContMixedState}

\end{document}